\pdfoutput=1
\documentclass[aps,twocolumn]{revtex4}

\usepackage{amsfonts}
\usepackage{amsmath}
\usepackage{amssymb}
\usepackage{version}
\usepackage{graphicx}

\pdfoutput=1
\includeversion{New_connection}
\excludeversion{Old_connection}

\begin{document}

\preprint{}
\title[Reassessment of MQT]{Investigation of low temperature quantum
crossover in Josephson junctions}
\author{James A. Blackburn$^{1}$, Matteo Cirillo$^{2}$, Niels Gr\o %
nbech-Jensen$^{3}$}
\affiliation{$^{1)}$Physics \& Computer Science, Wilfrid Laurier University, Waterloo,
Ontario N2L 3C5, Canada}
\affiliation{$^{2)}$Dipartimento di Fisica and MINAS-Lab, Universit\`{a} di Roma
\textquotedblleft Tor Vergata\textquotedblright\ I-00133 Roma, Italy}
\affiliation{$^{3)}$Department of Mathematics and Department of Mechanical and Aerospace
Engineering, University of California, Davis, CA 95616, USA}

\begin{abstract}
The evidence for macroscopic quantum tunneling (MQT) in Josephson junctions
at low temperatures has been reassessed. \ Swept bias escape distributions
have been modeled with an algorithm-based simulation and the results
compared with data from representative published experiments. Signatures
expected of a crossover to MQT are not found in the analyzed data.
\end{abstract}

\maketitle

\section{Introduction}

The physics of Josephson junctions, for a number of fundamental purposes and
device applications, can be described by a classical circuit model and
related dynamics \cite{BaroneVanD}. At the lowest temperatures, however,
there is another view in which the junction is presumed to transform into a
macroscopic quantum device, a \textquotedblleft qubit\textquotedblright\ , thereby acquiring properties specific
of quantum states such as tunneling and entanglement \cite{Leggett2002}.

We have previously provided evidence that the non-quantum Resistively and
Capacitively Shunted Junction (RCSJ) model of a Josephson junction can
successfully replicate a number of experiments originally interpreted as
manifestations of a macroscopic quantum state. These effects include
microwave induced transitions, Rabi oscillations, Ramsey fringes, and
entanglement of superconducting qubits \cite{Blackburn1}.\ 

It is worth noting that new classes of superconductive devices and systems 
relying on the physics of Josephson tunneling have recently attracted the attention of the applied 
superconductivity community. In particular, new concepts in radiation detection have 
appeared in the literature \cite{Oelsner1, Oelsner2}. This paper presents an analysis of thermal and 
quantum effects associated with the washboard potential of the Josephson effect. 
Besides the specific analysis involving thermal and quantum excitations, our work might 
bring deeper insight into Josephson potentials and their escape mechanisms, which could 
be relevant in applications. 

Here we present a more specific and systematic investigation
of the basic phenomena from which the new \textquotedblleft qubit\textquotedblright\ physics took
motivation.The earliest indication of Macroscopic Quantum Tunneling (MQT)
was provided by the least complicated experimental protocol: The bias
current to a Josephson junction is applied in a steadily increasing sweep
until the junction abruptly switches to a finite voltage state.\ The value
of the current at that moment is recorded and the sweep is restarted many
times. \ The accumulated data form a so-called switching current
distribution (SCD) that for any specific junction temperature is peaked at a
particular value of bias. The first claims of evidence of the macroscopic
quantum phenomenon came from this kind of experiment \cite{Voss}, i.e.
evidence of MQT; similar claims by other authors have been reported since
the mid-80's \cite{Devoret}.

It has been the standard approach in published work to infer the escape rate
out of the Josephson potential \cite{Blackburn1} from the observed SCD
distributions, and then to compare that \ rate with theory. Here we do just
the opposite: we will infer the SCD peaks from an appropriate expression for
the escape rate and then compare those predictions with experiment. We do
this with a simulation algorithm that has been previously discussed \cite%
{Blackburn2012}. Any theoretical model can thus be tested by judging if it
can duplicate experimental results.

Previously, we employed data from the experiment of Yu et al. \cite{Yu} as a
basis of assessing the predictions of a non-quantum device model from SCD
peaks. Here we again use that same set of data, as well as results in the
seminal work of Voss and Webb \cite{Voss}, but now exclusively from the
perspective of macroscopic quantum device models. We also consider some data
of the same type measured for junctions made of high temperature
superconductors. This allows a side by side comparison of non-quantum and
quantum predictions.

For the purposes of this treatment we suppose that the measured bath
temperature is indeed the same as the junction temperature.

\section{Theory}

The RCSJ model of a Josephson junction consists of three parallel elements:
a shunt resistor R, a shunt capacitor C, and a pure Josephson element. It
has a long history of successfully predicting the dynamics of Josephson
systems.

The current through a Josephson element is given by $I_{C}\sin\varphi$ and
the voltage across the element is governed by $d\varphi/dt=2eV/\hbar$, where 
$\varphi$ is the phase of the junction and $I_{C}$ is its critical current.
With a total applied bias current $I$, the phase dynamics is governed by

\begin{equation}
\frac{\hbar C}{2e}\frac{d^{2}\varphi}{dt^{2}}+\frac{\hbar}{2eR}\frac{%
d\varphi }{dt}+I_{C}\sin\varphi=I  \label{Eq1}
\end{equation}

If time is normalized to $1/\omega_{J0}$ where $\omega_{J0}=\sqrt {%
2eI_{C}/\hbar C}$ is the zero-bias Josephson plasma frequency, then

\begin{equation}
\ddot{\varphi}+\alpha\dot{\varphi}+\sin\varphi=\eta  \label{Eq2}
\end{equation}
where $\eta=I/I_{C}$ is a normalized bias current and $\alpha=\hbar\omega
_{J0}/2eI_{C}R$ is a normalized loss coefficient.

A Josephson junction with phase $\varphi$ has stored potential energy $%
E_{J}(1-\cos\varphi)$. The pre-factor in this expression is the Josephson
energy:

\begin{equation}
E_{J}=\hbar I_{C}/2e  \label{Eq3}
\end{equation}

The total potential energy of a junction, when an additional bias current is
supplied, is

\begin{equation}
U=E_{J}\left\{ \left( 1-\cos\varphi\right) -\eta\varphi\right\}  \label{Eq4}
\end{equation}

In this form it is apparent that the phase dynamics can be viewed in terms
of a fictitious `particle' moving in a potential $U$.

At zero bias the potential is a horizontal washboard and the `particle'
would sit at the bottom of the well at $\varphi=0$. \ Small oscillations
around the minimum of that well occur at the plasma frequency $%
f_{J0}=\omega_{J0}/2\pi$. At non-zero values of the bias: (a) the washboard
tilts (b) the minimum of the well occupied by the particle shifts to values $%
\varphi>0$ (c) the wells in the washboard potential become progressively
shallower, with correspondingly smaller plasma frequencies, $%
f_{J}=f_{J0}\left( 1-\eta^{2}\right) ^{1/4}$ and reduced barrier heights $%
\Delta U=2E_{J}\left[ \sqrt{\left( 1-\eta ^{2}\right) }-\eta\cos^{-1}\eta%
\right] $ \cite{BlackburnPLA} that both disappear at a bias equal to the
junction critical current.

The first escape mechanism to be recognized was classical thermal activation
(TA) in which the \textquotedblleft particle\textquotedblright\ jumps over
the barrier and then bounces down the washboard generating a voltage. When
the bias current is ramped up, a switching event is signalled when the
'particle' escapes from the well and a voltage appears across the junction.
At all but the lowest temperatures, the mechanism of this escape is thermal
activation over the barrier and into the running state. The escape rate for
such a process is \cite{Kramers}

\begin{equation}
\Gamma_{TA}=f_{J}\exp\left( \frac{-\Delta U}{k_{B}T}\right)  \label{GAMMATA}
\end{equation}
$k_{B}$ is Boltzmann's constant, and $T$ is the junction temperature.

It has been proposed that for sufficiently small temperatures, the junction
will change from a classical entity to a macroscopic quantum entity in which
case escape would occur via quantum tunneling. \ The escape rate for this
process is given by (see e.g. \cite{Devoret, Martinis}).

\begin{equation}
\begin{array}{l}
{\Gamma_{MQT}=a_{q}\;f_{J}\exp\left[ -7.2\frac{\Delta U}{hf_{J}}\left( 1+%
\frac{0.87}{Q}\right) \right] } \\ 
{a_{q}\approx\left[ 120\pi\left( \frac{7.2\Delta U}{hf_{J}}\right) \right]
^{1/2}}%
\end{array}
\label{Eq18}
\end{equation}
where $Q$ is the quality factor of the junction ($Q=\omega_{J}RC$).

The expression for escape due to quantum tunneling, Eq.(\ref{Eq18}), is a
limiting form generally considered applicable only for the lowest
temperatures. The question of possible finite temperature enhancements to
this rate was explored in a number of theoretical papers \cite{Grabert,
Martinis2}. \ It is a widely held opinion that any judgment as to whether
quantum theory actually does describe experiments must include temperature
enhancement effects in a revised expression for the escape rate. MQT escape
rates are expected to increase with finite temperature through thermal
enhancement \cite{Washburn, Grabert, Martinis2,Cleland}. We note that an
enhanced escape rate (greater than $\Gamma_{MQT}$) means it is easier to
escape from the well, so escape will occur sooner in the sweep. Therefore a
finite temperature effect will result in SCD peaks being shifted to lower
bias positions. For this reason the MQT escape peak expected from Eq.(\ref%
{Eq18}) must represent \ a \textit{maximum} possible bias position - no
quantum peak can advance beyond this point no matter how low the sample
temperature. \ So there is a \textquotedblleft cutoff\textquotedblright\
value for activation peak positions.

This finite temperature effect, as it applies to the particular case of
Josephson junctions, appeared in \cite{Martinis2} where the enhanced escape
rate in the weak damping limit was obtained from the zero temperature rate
Eq.(\ref{Eq18}) according to the following equations (Eqs.3.16, 3.3, 3.11 in 
\cite{Martinis2}):

\begin{equation}
\ln\left[ \Gamma(T)/\Gamma(0)\right] =10\pi\alpha(B-\frac{8}{5}%
)(k_{B}T/hf_{J})^{2}  \label{Martinis}
\end{equation}
where:

\begin{align}
B & =(\Delta U/\hbar\omega_{J})s(\alpha)  \label{parameter1} \\
s(\alpha) & =\frac{36}{5}\left[ 1+\frac{45}{\pi^{3}}\xi(3)\alpha\right]
\label{parameter2}
\end{align}
with a damping constant $\alpha=1/2Q$ for the Josephson junction and $%
\xi(3)=1.202$ is a Riemann number.

\section{Simulations}

An algorithm for computer simulations of swept bias experiments was
described in \cite{Blackburn2012}. The program is built around appropriate
escape rate expressions Eq.(\ref{GAMMATA}) and Eq.(\ref{Eq18}) or Eq.(\ref%
{Martinis}), and requires values for the following input parameters:
junction critical current $I_{C}$, junction capacitance $C$, the junction
quality factor $Q$, and the time taken for a bias sweep. \ There are \textbf{%
no} adjustable parameters.

\begin{figure}[tbp]
\begin{center}
\includegraphics[
height=4.3526in,
width=3.3814in
]{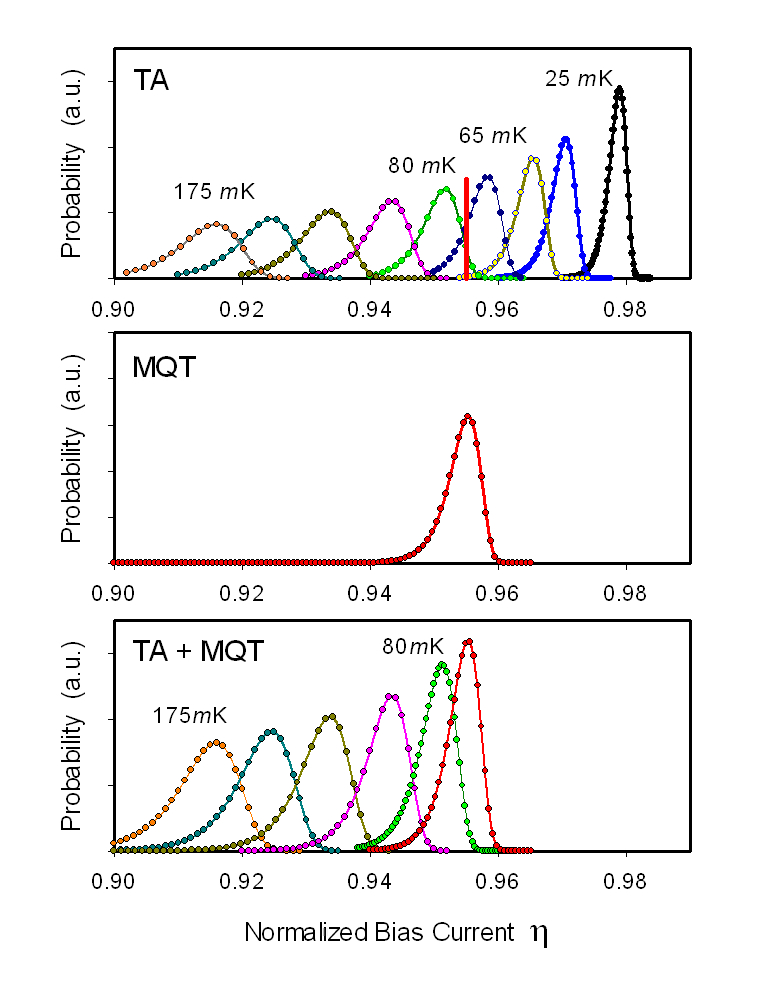}
\end{center}
\caption{Switching current distributions generated by algorithm-based
computer simulations of a swept bias experiment. Bias currents are
normalized to a junction critical current $I_{C}=1.957\;\protect\mu A$. \
Upper panel: SCD peaks from just thermal activation at various temperatures
(the temperatures of unlabeled peaks are $150,$ $125,100$, then $50,$ $40$ $%
mK$). Middle panel: solid (red) line - MQT peak for $Q=12$ . Lower panel:
simulations of SCD peaks when escape rates from both thermal activation and
quantum tunneling coexist. Note that the peaks `freeze' at the location of
the MQT peak.}
\label{TAMQT}
\end{figure}

For these simulations we chose to model a Josephson junction with the
following parameters taken from \cite{Yu}: bias ramp time $4.89$ \textit{ms}%
, $I_{C}=1.957\;\mu A$, $C=620\;fF$, and $R=300\;\Omega $.\ The zero bias
plasma frequency was thus $15.59$ GHz. \ The simulations reported here
represented an equivalent of $1,000,000$ repeated bias sweeps. Sample
results are depicted in Fig.\ref{TAMQT}. 

On the upper panel (TA) the SCD peaks from the simulation of escape only by
thermal activation are presented. On the center panel (MQT), the single SCD
peak from macroscopic quantum tunneling alone with an assumed $Q=12$ is
shown. Note that the MQT result (red peak) corresponds to a TA peak at $%
T\approx70$ \textit{mK}.

Next we modified the original simulation algorithm to include both the
thermal activation rate Eq.(\ref{GAMMATA}) and the quantum tunneling rate
Eq.(\ref{Eq18}). That is, TA \textit{and} MQT are present throughout the
swept bias procedure. A simulation run produced the results shown in the
lower panel (TA+MQT) of Fig.\ref{TAMQT}. Here it is evident that when both
escape modes are running concurrently, TA peaks for high enough temperatures
appear at the same positions that they held in the absence of an MQT escape
mode, but below this temperature only the MQT peak is seen.

This behavior is due to the nature of SCD peaks -- they are distributions of
the probability that an escape will occur at a particular bias value\ during
a sweep. These peaks exhibit the property that on the high side of the
maximum the probability returns to zero. A second process with a lower
escape rate would have its SCD peak at a higher bias, but that bias region
cannot be accessed. Therefore swept bias experiments will reflect only the
process with the higher escape rate.

\begin{figure}[tbp]
\begin{center}
\includegraphics[
height=2.2in,
width=3.1669in
]{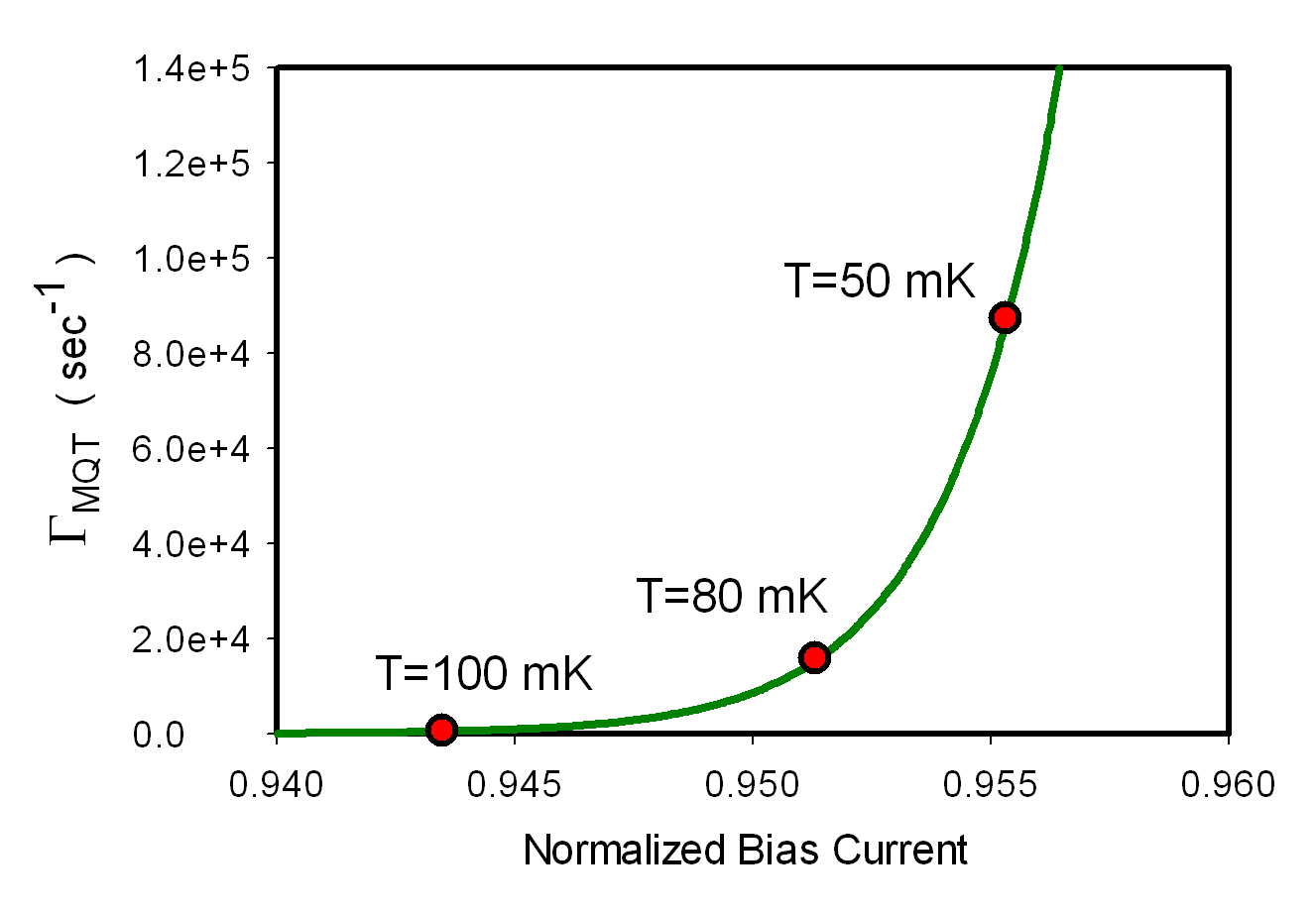}
\end{center}
\caption{Macroscopic quantum tunneling escape rate as
a function of bias current (solid line) with superimposed markers at the
positions of the lowest temperature peaks in the lower panel Fig.\protect\ref%
{TAMQT}.}
\label{MQT_GAMMA}
\end{figure}

Note that the Josephson plasma frequency is bias-dependent because the
curvature of the well is also bias-dependent. For example, in the harmonic
approximation, $\ f_{J}=f_{J0}\left( 1-\eta ^{2}\right) ^{1/4}$. In
addition, the barrier height $\Delta U=2E_{J}\left[ \sqrt{\left( 1-\eta
^{2}\right) }-\eta \cos ^{-1}\eta \right] $ is also bias-dependent. The
thermal activation rate Eq.(\ref{GAMMATA}) and the thermally enhanced MQT rate %
Eq.(\ref{Martinis}) both explicitly depend on temperature, but the MQT escape
rate Eq.(\ref{Eq18}) does not. However, the MQT rate will vary during the process of
sweeping the bias current. This is shown in Fig.\ref{MQT_GAMMA}.
What this means is that the MQT escape rate cannot be regarded as a constant; its value
depends on where in a sweep the SCD peak is located.

\section{Crossover}

The crossover temperature is defined by the condition that the escape rate
for thermal activation equals the escape rate for MQT; hence for Eqs.(\ref%
{GAMMATA},\ref{Eq18}),%
\begin{equation}
f_{J}\exp\left( \frac{-\Delta U}{k_{B}T}\right) =a_{q}\;f_{J}\exp\left[ -7.2%
\frac{\Delta U}{hf_{J}}\left( 1+\frac{0.87}{Q}\right) \right]  \label{Eq19}
\end{equation}
With the usual assumption $Q\gg1$ this leads to

\begin{equation}
T_{cr}\approx \frac{hf_{J}}{2\pi k_{B}}  \label{Eq21}
\end{equation}%
This is a widely quoted result \cite{Devoret,Martinis,Voss,Wallraff,
Washburn}.

Because the escape rates for TA and MQT are equal at this temperature, it is
considered to be the point at which there is a crossover between the two
escape modes. It should be noted that the rate expression Eq (\ref{Eq18}) is
in fact a \textbf{zero temperature limit} \cite{Devoret, Martinis}.

\begin{figure}[tbp]
\begin{center}
\includegraphics[
height=3.3235in,
width=3.2984in
]{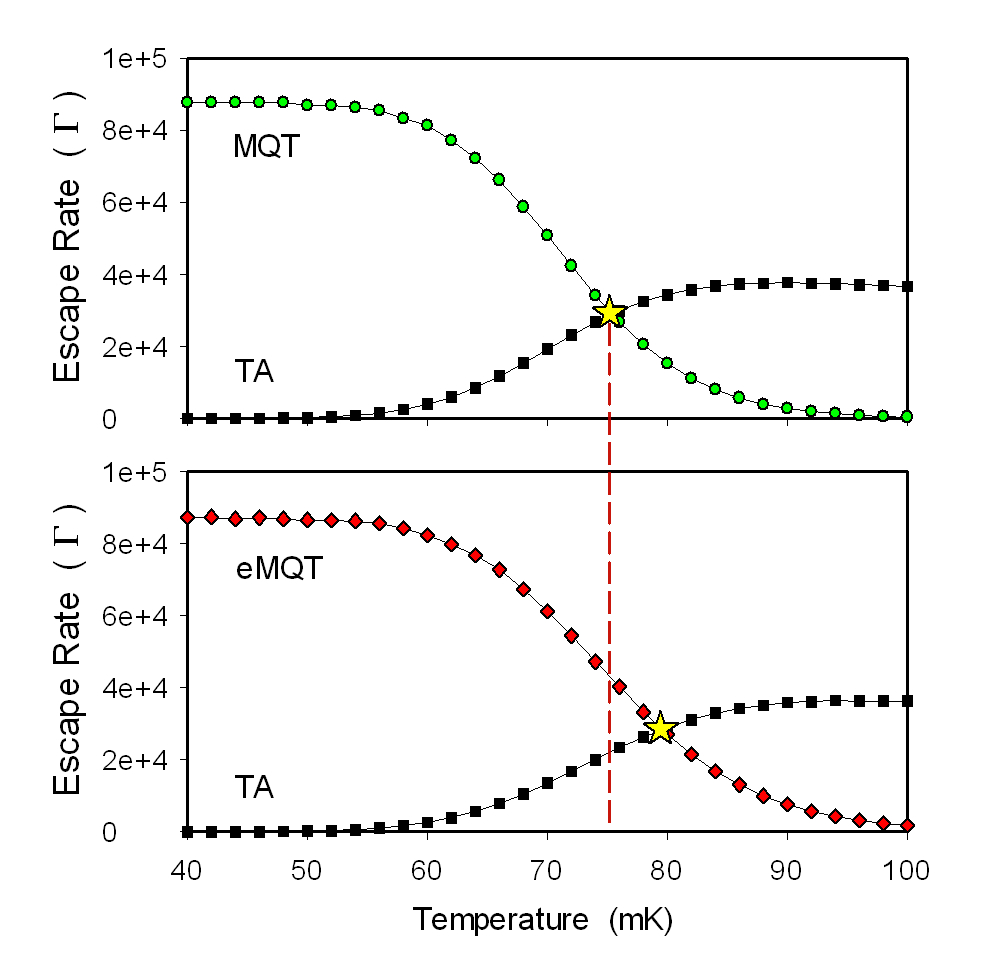}
\end{center}
\caption{Crossover in the escape rates due to thermal activation and
macroscopic quantum tunneling. Upper panel: MQT rate from Eq.(\protect\ref%
{Eq18}) and TA rate from Eq.(\protect\ref{GAMMATA}). Note that the temperature dependence of MQT 
must be interpreted as illustrated in Fig.\ref{MQT_GAMMA}. \ The star marks the
crossover point at $75.2mK$. Lower panel: enhanced MQT rate from Eq.(\protect
\ref{Martinis}) and TA rate from Eq.(\protect\ref{GAMMATA}). \ The star
marks the crossover point $79.5mK$ for this enhanced tunneling process.}
\label{crossovers}
\end{figure}

This changeover to thermal activation, from both macroscopic quantum
tunneling and for thermally-enhanced macroscopic tunneling, is illustrated
in Fig.\ref{crossovers}. 

In the simulation, for a chosen temperature, the bias is ramped up until an
escape peak is reached. \ The bias current at that point determines the
potential barrier $\Delta U$ and also the plasma frequency. Therefore the thermal escape rate can be
calculated from Eq.(\ref{GAMMATA}). \ Likewise the MQT escape rate can be
calculated from Eq.(\ref{Eq18}). In the lower panel, this process has been
repeated but with the thermally enhanced escape rate given by Eq.(\ref%
{Martinis}).\ What this shows is that for an MQT process, there is a
crossover temperature $T_{cr}$, whereas for an enhanced MQT process there is
a corresponding \textit{enhanced} crossover temperature $eT_{cr}$. \ As the
figure illustrates, this enhancement of the \ crossover temperature can be
quite small, $\approx 4.5mK$.

\section{Comparison with Experiments}

We now analyze the experiments presented in \cite{Yu}. \ Figure 2 in \cite%
{Yu} (also repeated in \cite{Yu2}) includes a plot of SCD peak positions as
a function of junction temperature. As discussed in the Appendix of \cite%
{Blackburn1}, peak positions are a more precise indication of behavior at
the lowest temperatures where the width of the peaks shrinks considerably.

\begin{figure}[ptb]
\begin{center}
\includegraphics[
height=2.4633in,
width=3.4637in
]{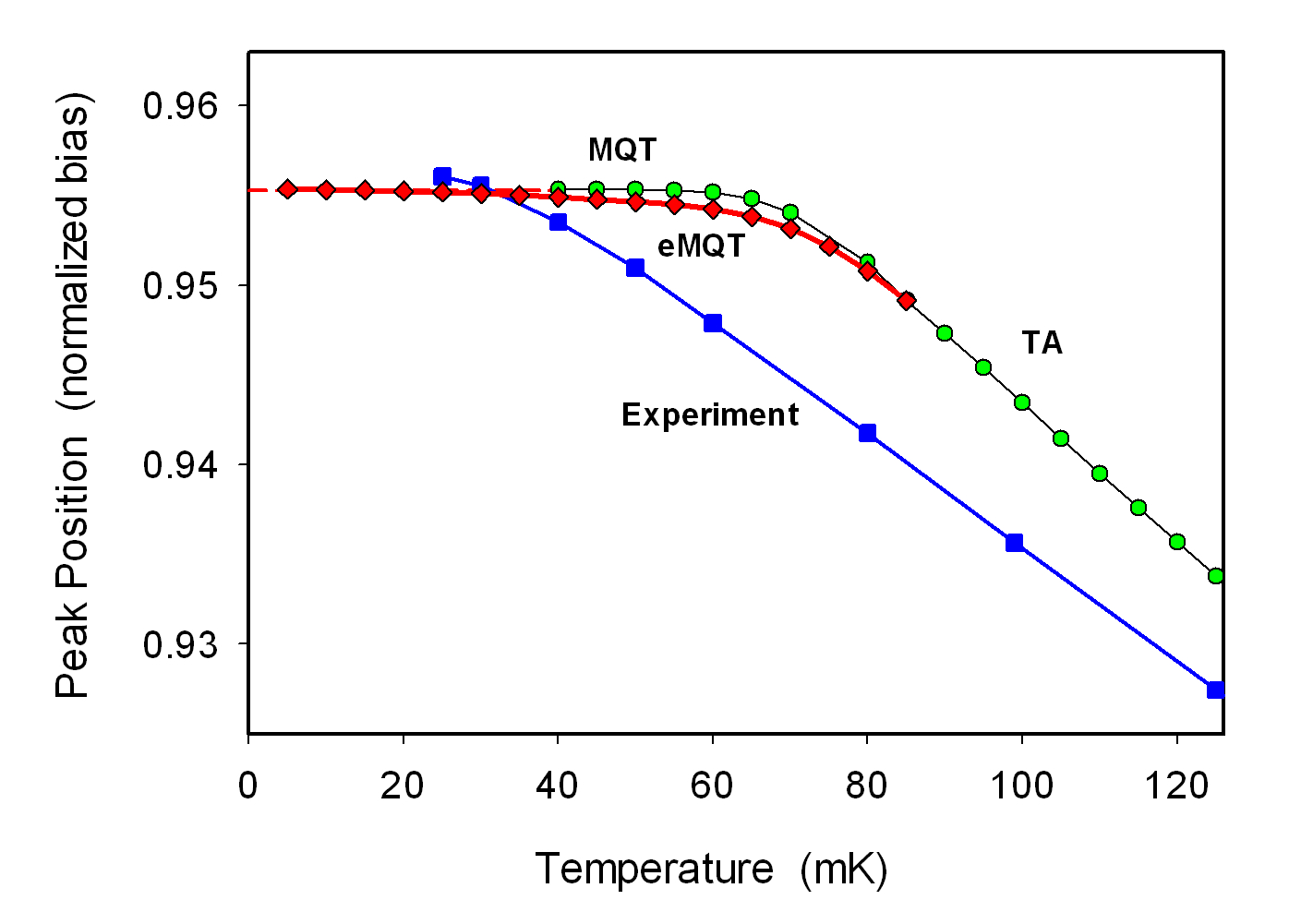}
\end{center}
\caption{Squares (blue): experimental data for SCD peak positions from Fig.2
in \protect\cite{Yu}. Circles (green): simulation results for SCD peak
positions versus temperature with TA and MQT operating concurrently (as in
the lower panel of Figure \protect\ref{TAMQT} ). \ Diamonds (red):
simulation using an escape rate that includes thermal enhancement of
tunneling (eMQT) from Eq.(\protect\ref{Martinis}). In the simulation plots,
branches are labeled according to the dominant escape processes: TA, MQT,
 and eMQT.}
\label{theoryanddata}
\end{figure}

We digitized the eight lowest temperature data points and replot them as
(blue) squares in Fig.\ref{theoryanddata}. Note that the temperature scale
here is linear. 

From simulation-generated plots such as shown in the lower panel of Fig.\ref%
{TAMQT} with TA and MQT processes acting concurrently, peak positions were
extracted. The solid circles (green) in Fig.\ref{theoryanddata}\ come from
that process. This illustrates the natural transition that takes place
between thermal activation and quantum tunneling.

It might be observed that the experimental data lie below the TA simulation
results. However, at temperatures above $\approx200$ \textit{m}K simulation
and experiment are in excellent agreement, as is apparent in Fig.2 of \cite%
{BlackburnEPL}. In \cite{BlackburnEPL} we speculated on why experimental
results might peel-away from the expected TA behavior at low temperatures.

We applied Eq.(\ref{Martinis}) for the enhanced escape rate applicable to
this experiment with parameter values $\alpha=1/2Q$ and $Q=12$, $T_{0}=65$ 
\textit{mK}, $\omega_{0}/2\pi=15.59GHz$. \ The simulation yielded the results
plotted as diamonds (red) in Fig.\ref{theoryanddata}. As would be
anticipated, the predicted peak positions with thermal enhancement
(diamonds) lie slightly below the corresponding simulation results without
enhancement (circles). \ As expected from the simulation results shown in
Fig.\ref{crossovers}, there has been a small upward shift in the crossover
point.

With respect to \textquotedblleft error bars\textquotedblright\ that might be associated with the data
presented in Fig.\ref{theoryanddata}, we note the following.

For the \textit{experiments}, escape data from each bias scan were sorted
into bins of width $1nA$. The large number of repetitive scans ($%
5\times10^{4}$ in this instance) assured that the bin with the largest
number of counts corresponded to the true position of the SCD peak with a
statistical uncertainty of $\pm$ one bin width. The junction critical
current was $1.957\mu A;$ hence the uncertainty bars for these experimental
data points would be $0.001/1.957\approx0.0005$, in normalized units. In Fig.%
\ref{theoryanddata} the experimental data are shown as (blue) squares which
have a vertical dimension of $0.001$; consequently for the experiments,
error bars would be approximately one half the height of the square symbols.

For the \textit{simulations} based on escape rates, there were $50,000$ data
bins evenly spaced over the normalized bias interval $0\rightarrow1.0$. Each
simulation run represented an equivalent of $10^{6}$ repeated bias sweeps.
The normalized bin spacing is $1.0/50,000=0.00002$, which is less than a
tenth of the uncertainty of the experimental data, and also much smaller
than the thickness of the solid line (red) that extends towards $T=0$. Hence
statistical error bars on simulation results would be too small to be
visible. Therefore the disagreement between experiment and quantum
predictions cannot be attributed to statistical errors.

\begin{figure}[tbp]
\begin{center}
\includegraphics[
height=2.4327in,
width=3.1669in
]{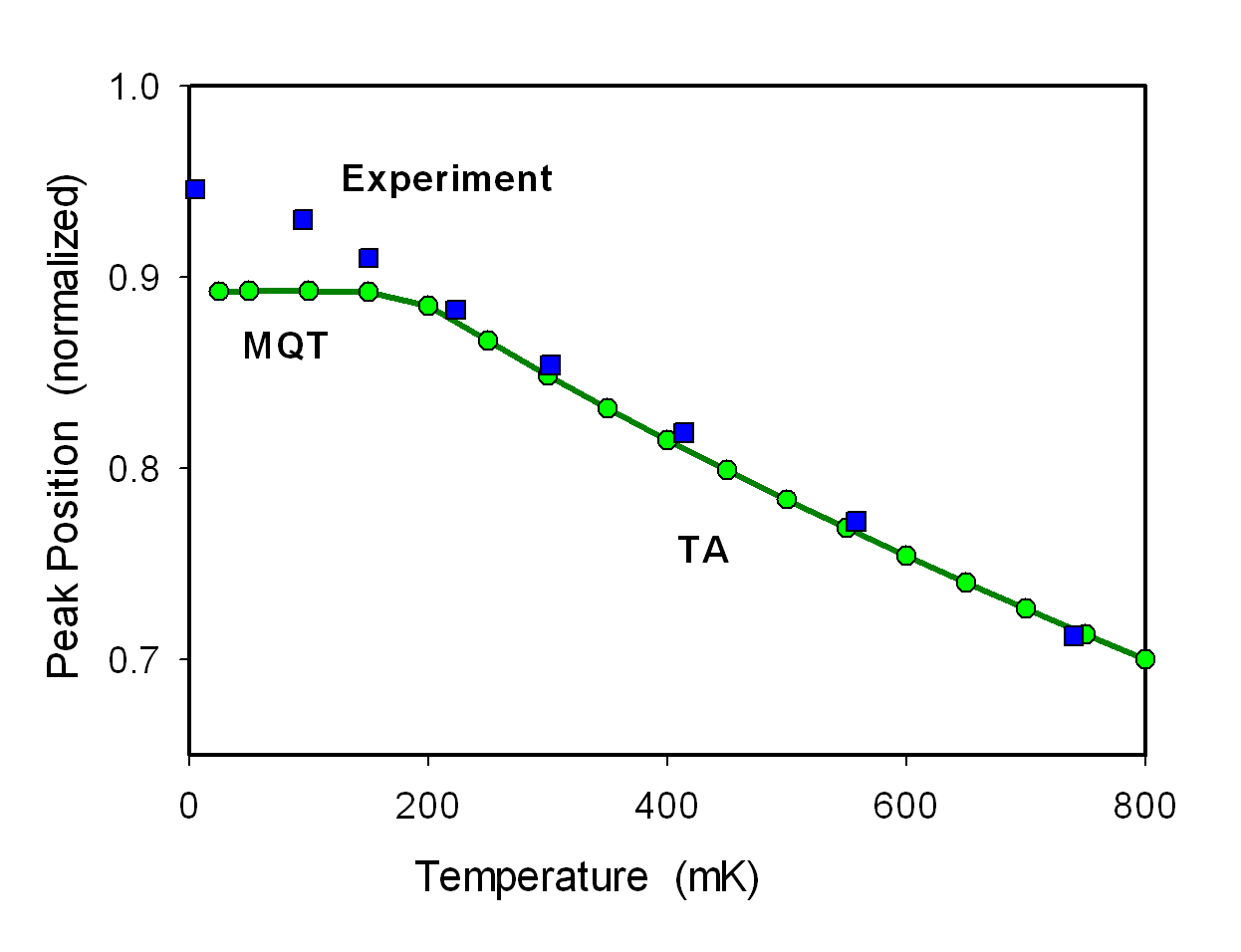}
\end{center}
\caption{Squares (blue): experimental data extracted from Fig.1 in 
\protect\cite{Voss}. \ Circles (green): simulation results for SCD peak
positions versus temperature with TA and MQT operating concurrently.}
\label{VossWebb}
\end{figure}

We performed the same type of analysis done for the Yu et al. experiment 
\cite{Yu} on the data reported by Voss and Webb \cite{Voss}. The parameters
for this simulation were: $I_{C}=1.62\mu A$, $C=0.1pF$, $Q=7.1$, zero bias
plasma frequency $35.3GHz$, and a bias ramp time of $0.01s$. The result of
the analysis is shown in Fig.\ref{VossWebb}. 
Error bars associated with both experimental and simulation data points in
this figure can be estimated in the same manner as was done in connection
with Fig.\ref{theoryanddata} and the results are, as before, that on the
scale of this plot the uncertainties are insignificant.

Two significant features can be seen in Fig.\ref{VossWebb}. The first is
that the experimental data show no evidence of a quantum transition since
the peak positions continue to move up as the temperature of the junction
decreases below the theoretical crossover temperature. \ 

Secondly, the experimental data points closely track the simulation results
in the "thermal activation" regime of bath temperatures. We note that in Fig.%
\ref{theoryanddata} the experimental data do not follow the thermal bath
temperature, at least for the range of temperatures in the plot. This
phenomenon has been previously interpreted as evidence that the sample
temperature might be different from the bath temperature \cite{BlackburnEPL}%
. The fact that in the experiment of Voss and Webb the temperature of the
sample is identical to that of the bath could be due to their declared
experimental condition that the sample was mounted inside the mixing chamber
of the dilution refrigerator and not externally anchored to the chamber
itself (as in most other experiments).

\begin{figure}[ptb]
\begin{center}
\includegraphics[
height=2.4327in,
width=3.1669in
]{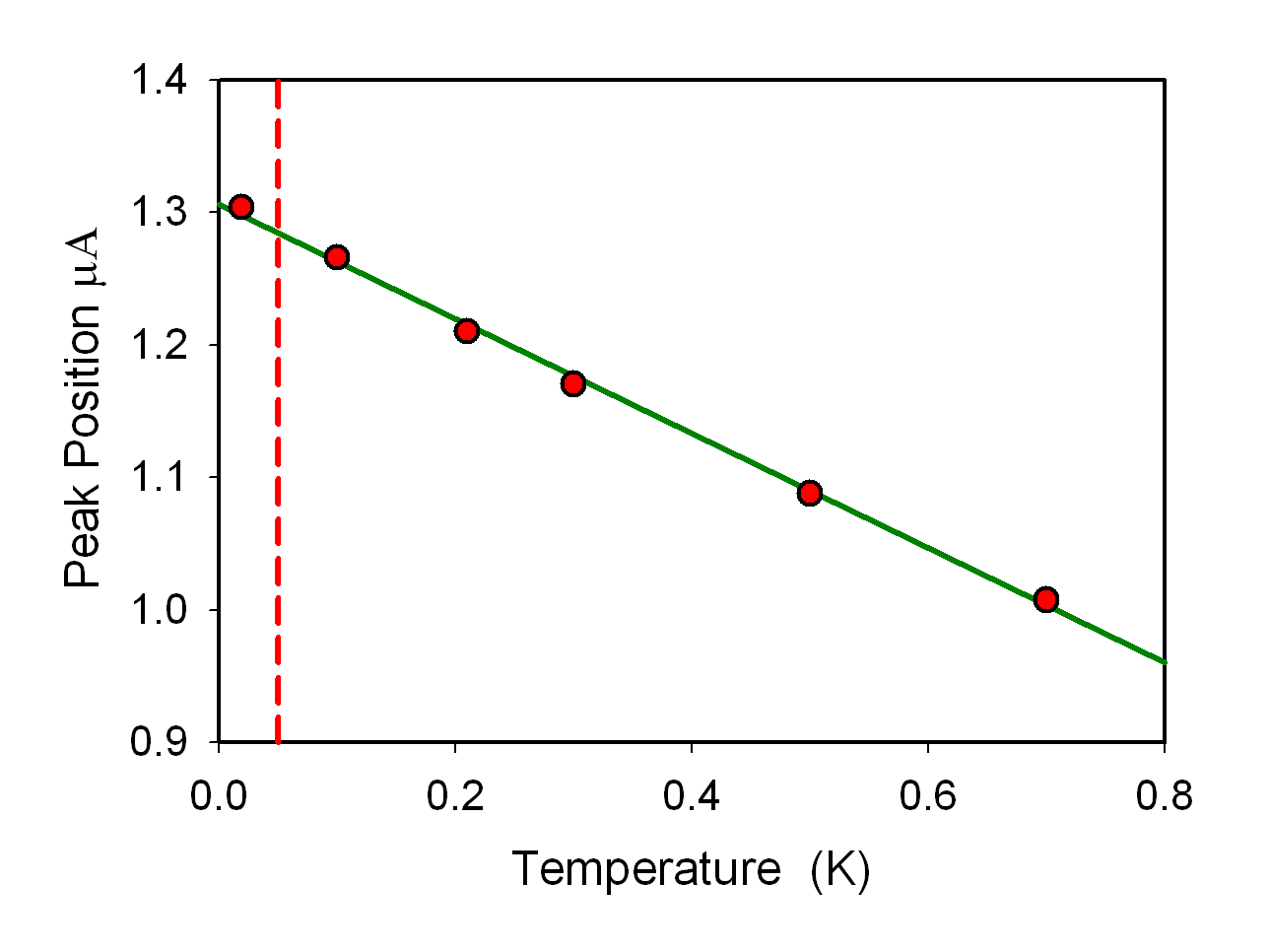}
\end{center}
\caption{Experimental data extracted from Fig, 2(a) in \protect\cite{Bauch}
showing SCD peak positions as a function of temperature for a high-$T_{C}$
Josephson junction. \ The dashed line marks the reported crossover
temperature of $50mK$ and the solid straight line is a guide to the eye to
emphasize the lack of any special behavior below the crossover point. \ The
junction critical current was $I_{CO}=1.40\protect\mu A$.}
\label{Bauch}
\end{figure}

In Fig. \ref{Bauch} we show data digitized from plots of an experiment
performed on high-Tc superconducting materials, namely a grain boundary
biepitaxial junction \cite{Bauch}. In the figure we display the experimentally determined position of the
switching current peaks as a function of temperature. Although we have not
performed a direct simulation in this case, due to the lack of sweep rate
data in the paper, the nearly linear behaviour down to very low temperatures
is so striking that we can very reasonably claim that no transition to the
theoretically predicted MQT curve has occurred.

\section{Comments}

As mentioned already, there exist other sets of experiments in the
literature reporting SCD experiments with claims of evidence for MQT. For
this study we have chosen to focus on certain experiments that are
sufficiently straightforward for us to suppose there are no unknown and/or
uncontrollable effects that might disturb the nature of the mechanism by
which escape takes place out of the one-dimensional Josephson potential. Low
supercurrent junctions represent somewhat `safer' candidates from this point
of view. It has been shown, for example, that the nature of the escape
processes in high-Tc junctions can be unusual \cite{Barbanera} since this
kind of tunnelling structure can exhibit phase-gradient effects for
relatively high current densities.

A remark in \cite{Devoret} captures an early view of the crossover process:
\textquotedblleft The crossover temperature at which the escape rate changes
from thermal (temperature dependent) to quantum (temperature independent) is
predicted to be $\hbar\omega_{p}/2\pi k_{B}$ in the limit $Q\gg1$%
\textquotedblright. \ That picture has evolved somewhat. \ Figure 3 in \cite%
{Grabert} summarizes the current suppositions underlying the macroscopic
quantum tunneling hypothesis. It distinguishes between various temperature
intervals: well below the crossover region, quantum tunneling prevails;
within the crossover interval itself, quantum tunneling and thermal
activation both operate; above the crossover zone quantum corrections apply
to the classical thermal escape process; and then finally beyond that,
thermal hopping as the only escape mode.

The present simulations are consistent with this picture. With respect to
switching current distributions, a transition in SCD peak behavior is
expected even when both processes coexist, and may indeed coexist\ on either
side of the crossover point. It follows that any evidence of freezing of
escape peak positions at temperatures below the crossover point would give
no information about when the macroscopic quantum state actually coalesced
-- one could only conclude that it must have been formed at some temperature
above $T=T_{cr}$. \ Even below the crossover temperature a Josephson
junction could not be supposed to be fully quantum. \ This could have
consequences for the operation of qubits.

In \cite{Washburn} Eq.(\ref{Martinis}) appears in a slightly altered form,
with the right hand side expressed simply as $s(\alpha)T^{2}$. This
emphasized the expectation of a $T^{2}$ dependence of the enhanced escape
rate near $T=0$. With that in mind, Washburn et al. inferred escape rates
from experimental SCD peak data, and then plotted the natural logarithm of
that escape rate vs $T^{2}$. From that plot, the slope $s(\alpha)$ was
extracted. Tellingly, they stated \textquotedblleft we do not find
quantitative agreement with theoretical prediction for the slope $s(\alpha)$%
\textquotedblright\ and in the summary concluded that they achieved only
qualitative agreement with theoretical predictions. \ Certainly, this fell
short of confirming thermal enhancement of MQT. \ Similar outcomes were
reported in \cite{Cleland}.

In summary, for the first time escape rates arising from macroscopic quantum
tunneling theory have been tested against experiments. We have extracted
switching current distribution data from three selected experiments and
compared those data with simulation-based predictions of both zero
temperature MQT theory, and thermally enhanced MQT theory. We emphasize that
no comparisons of this type have been reported before. Significantly, the
scaling expression for the anticipated thermal enhancement of escape rates
in a Josephson junction with small damping does not resolve discrepancies
between zero temperature theory and observation, at least for the selected
experiments.

\section{Acknowledgement}

We are deeply indebted to Hermann Grabert for clarifying some key concepts
relating to thermal enhancement and for guiding us to the crucial scaling
relation that allowed escape rates to be determined in the region below the
crossover temperature

\end{document}